\documentclass[12pt]{article}
\def\be{\begin{equation}}
\def\ee{\end{equation}}
\def\beq{\begin{eqnarray}}
\def\eeq{\end{eqnarray}}
\def\n{\nonumber}

\begin{document}
\titlepage
\vspace*{2cm}
\begin{center}

{\Large \sc Lie symmetries for equations in
conformal geometries}
\vspace{1cm}

{\large \bf S Hansraj$^{\dag,a}$, S D Maharaj$^{\dag,b}$, A M Msomi$^\ddag$ and K S Govinder$^{\dag,c}$}
\vspace{5mm}

{\it $^\dag$ Astrophysics and Cosmology Research Unit,  School of Mathematical Sciences, University of KwaZulu--Natal,
Durban 4041, South Africa\\}
{\it $^\ddag$ Department of Mathematics, Technikon Mangosuthu, Durban 4000, South Africa\\}
{\it Email: $^a$hansrajs@ukzn.ac.za; $^b$maharaj@ukzn.ac.za; $^c$govinder@ukzn.ac.za \\}
\vspace{2cm}

\end{center}

\begin{abstract}
We seek exact solutions to the Einstein field equations which arise when two spacetime geometries
are conformally related. Whilst this is a simple method to generate new solutions to the field equations,
very few such examples have been found in practice. We use the method of Lie analysis of
differential equations to obtain new group invariant solutions to conformally related Petrov type D spacetimes.
Four cases arise depending on the
nature of the Lie symmetry generator. In  three cases we are in a
position to solve the master field equation in terms of elementary
functions. In the fourth case special solutions in terms of Bessel
functions are obtained. These solutions contain known models as special cases.
\end{abstract}

\newpage
\pagenumbering{arabic}
\section{Introduction}

Exact solutions, as opposed to computer generated solutions, of the Einstein field equations
are of immense importance in understanding the behaviour of a large variety of celestial
phenomena. The literature abounds with many different techniques that have been invoked in
an effort to obtain new exact solutions for different configurations of matter.

Conformal transformations, as a mathematical procedure, have also
been successfully utilised in obtaining new solutions. This is
amply illustrated by the comprehensive model of Castejon--Amenedo
and Coley (1992). Further, it is known that conformal structures
play an important role in twistor theory (Penrose 1999). The
restrictive feature of such analyses, however, is the complexity
and nonlinearity of the resultant field equations. Some
researchers have adopted the Newman-Penrose formalism which, on
account of severe integrability constraints arising out of the
Bianchi identities, have not proved fruitful in general.

An alternative is to assume the existence of a conformal symmetry on the
manifold in order to generate solutions to  the Einstein field
equations, as these symmetries impose additional restrictions on the
metric tensor field and the field equations may be simplified. The
physical significance of conformal Killing vectors is that they
generate constants of the motion along null geodesics for massless
particles.

The maximal spanning $G_{15}$
of conformal motions for  Minkowski space is given by Choquet--Bruhat
$ et$ $al$ (1982), and for Robertson--Walker spacetimes by
Maartens and Maharaj (1986). In addition the conformal geometries of the
$pp$-wave spacetimes (Maartens and Maharaj 1991), static spherically
symmetric spacetimes (Maartens $ et {\hspace{2mm}}  al$ 1995,1996),
Bianchi
I and V locally rotationally symmetric spacetimes (Moodley 1992) and
Stephani spacetimes (Moopanar 1993) have been completely determined.
Kramer (1990) was able to generate a class of
 metrics for rigidly rotating perfect fluids which admit a proper
conformal Killing vector. Additionally rigidly rotating perfect fluids
admitting two Killing vectors and a proper conformal Killing vector were
studied by Kramer and Carot (1991). Axisymmetric spacetimes, in the
general case of differential rotation,  were examined by Mars and
Senovilla (1993, 1994).

Several
new results, utilising this conformal symmetry approach have recently been obtained by
Castejon--Amenedo and Coley (1992), Coley and Tupper (1990a,b,c), Dyer
$ et {\hspace{2mm}} al$ (1987) and Maharaj $ et$ $ al$
(1991). In particular, Maartens and
Mellin (1996) used the conformal symmetries of Bianchi I spacetime to
demonstrate that the expansion of anisotropic radiation universes tends
towards isotropy at late times.

Our interest lies  in spacetimes
that admit an $s$-dimensional Lie algebra $C_s$ of conformal motions.
We exploit the Defrise--Carter theorem (1975) to generate new models of perfect
fluids. We select a spacetime of Petrov type D and its conformally related counterpart in an effort to obtain new solutions to the associated Einstein field equations. We analyse the field equations and show that they can be reduced
to a simpler form.  The results of Castejon--Amenedo and
Coley are regained as a special case of
a more general class of exact solutions. Group invariant solutions to a particular field equation which acts as a
master equation for the entire system are sought.  This analysis generates a rich class of solutions. A number of  cases arise depending on
the nature of the Lie symmetry generator. In  all cases we are in a
position to provide solutions to the master field equation in terms of elementary
functions. This analysis demonstrates the value of the Lie
analysis of differential equations in this application.

\section{Spacetime Geometry}

We consider the line element \be ds^2 = -dt^2 + dx^2 +
e^{2\nu(y,z)} (dy^2 + dz^2)  \label{1} \ee which is of Petrov type
$D$. The spacetime (\ref{1}) admits three Killing vectors
\begin{eqnarray}
\label{1'}
{\bf X}_1 &=& \frac{\partial}{\partial t} \label{1'a}  \\ \n \\
{\bf X}_2 &=& \frac{\partial}{\partial x} \label{1'b} \\ \n \\
{\bf X}_3 &=& x\frac{\partial}{\partial t} + t \frac{\partial}{\partial x} \label{1'c}
\end{eqnarray}
which obey the commutation relations
\[
[{ \bf X}_1, {\bf X}_2] = 0 \hspace{15mm} [{\bf X}_1,{\bf X}_3]=0 \hspace{15mm}
[{\bf X}_2, {\bf X}_3] = 0
\]
The Lie algebra of Killing vectors is a $G_3$ of motions, with the group structure satisfying the
above relations.

Suppose that a manifold $(M, {\bf g})$ is neither conformally flat nor
conformally related to a generalised plane wave. Then according to a
theorem due to Defrise--Carter (1975),
 a Lie algebra of conformal Killing vectors
on $M$ with respect to ${\bf g}$ can be regarded as a Lie algebra of
Killing
vectors with regard to some metric on $M$ conformally related to ${\bf
g}$. Therefore if a spacetime admits the conformal
group  $C_s$,
then either it is conformally flat ($s = 15$), conformally related to a
generalised plane wave ($s \leq 7$), or the metric $\bar{g}_{ab} =
e^{2U}
g_{ab}$ where $g_{ab}$ admits an $s$-dimensional ($s\leq
6$) isometry group. The last possibility is evident above and so as our starting point we consider
the conformally related metric
\be
ds^2 = e^{2U}\left[-dt^2 + dx^2 + e^{2\nu(y,z)}\left( dy^2 +
dz^2 \right)\right] \label{2}
\ee
where $U = U(t, x, y, z)$.
The Killing vectors (\ref{1'}) are now conformal Killing vectors of the conformally related
spacetime (\ref{2}). The Weyl conformal tensor is given
by
\beq
3e^{2\nu}\bar{C}_{0101} = -6\bar{C}_{0202} = -6\bar{C}_{0303} &=&
6\bar{C}_{1212} \n \\ \n \\
= 6\bar{C}_{1313} = \frac{3}{5} e^{-2\nu}\bar{C}_{2323} =
\frac{3}{5}e^{-2\nu}\bar{C}_{3232}  &=& \nu_{yy} + \nu_{zz} \n
\eeq which clearly indicates that the metric (\ref{2}) is not
conformally flat in general. (We use the notation that overhead
bars on quantities are defined in the conformally related
spacetimes (\ref{2}).) Note that $\nu_{yy} + \nu_{zz} $ is not
zero.
 We make the assumption that
\be
\nu_{yy} + \nu_{zz} = -2ke^{2\nu} \label{2'}
\ee
where $k$ is a nonzero constant. This choice is made on the grounds of simplicity and follows
the treatment of Castejon--Amenedo and Coley (1992). In addition if $k=0$
then the line element (\ref{1}) becomes flat, and consequently
(\ref{2}) would be conformally flat. The condition (\ref{2'}), with
$k \neq 0$, obviates this occurrence.

To determine the perfect fluid energy--momentum tensor, we
select a fluid 4--velocity vector ${\bf u}$ that is noncomoving with the
form \be
u^a = e^{-U} \left( \cosh v \delta^a_0 + \sinh v \delta^a_1 \right)
\label{3}
\ee
where $v = v(t,x)$. Note that a  trivial calculation reveals that  for
$v = \mbox {constant}$, the perfect fluid Einstein field equations would
imply conformal flatness.

\section{Field Equations and Kinematics}

The Einstein field equations are given by
\begin{eqnarray} \label{212}
U_tU_y - U_{ty} &=& 0 \label{212a} \\ \n \\
U_tU_z - U_{tz} &=& 0 \label{212b} \\ \n \\
U_{x} U_{y} - U_{xy} &=& 0 \label{212c} \\ \n \\
U_x U_z - U_{xz} &=& 0\label{212d} \\ \n \\
U_{t}U_x - U_{tx} &=&
-\frac{1}{4} (\mu + p) e^{2U} \sinh 2v \n \\
 \label{212e} \\ \n \\
U_{y}U_{z} - U_{yz} + \nu_{z}U_{y} + \nu_{y}U_{z} &=& 0
\label{212f} \\ \n \\
 -2U_{xx} - U^2_x + 3U^2_t  \n \\ - e^{-2\nu} \left(
2U_{yy} + 2U_{zz} + U^2_y+ U^2_z  + \nu_{yy} + \nu_{zz} \right)
&=& (\mu + p)e^{2U}\cosh^2 v - pe^{2U}  \n \\
\label{212g}         \\ \n \\
-2U_{tt} -U^2_t + 3U^2_x \n \\ + e^{-2\nu} \left(
2U_{yy} + 2U_{zz} + U^2_y+ U^2_z  + \nu_{yy} + \nu_{zz} \right)
&=& (\mu + p) e^{2U}\sinh^2v  + pe^{2U}      \n \\
\label{212h} \\ \n \\
2U_{zz} + U^2_z + 3U^2_y + 2\nu_{y}U_{y} - \n \\
2\nu_{z}U_{z} + e^{2\nu} \left(2U_{xx} - 2U_{tt} + U^2_{x} - U^2_{t}
\right) &=& pe^{2\nu + 2U}   \label{212i} \\ \n \\
2U_{yy} + U^2_y + 3U^2_z - 2\nu_{y}U_{y} +
2\nu_{z}U_{z} \n \\ + e^{2\nu} \left(
2U_{xx} - 2U_{tt} + U^2_{x} - U^2_{t}
\right) &=& pe^{2\nu + 2U}  \label{212j}
\end{eqnarray}
for the line element $(\ref{2})$.

\vspace{5mm}

The vorticity ${\bar \omega}_{ab}$ is given by
\[
{\bar \omega}_{ab} = 0
\]
so that the gravitational field is irrotational. The components of the
acceleration $\bar{\dot{u}}^a$ are
\begin{eqnarray}\label{214}
\bar{\dot{u}}_0 &=& - \left[\left(v_x + U_t\right) \sinh^2v + \left(v_t +
U_x \right) \cosh v \sinh v \right] \label{214a} \\ \n \\
\bar{\dot{u}}_1 &=& \left(v_t + U_x\right) \cosh^2v + \left(v_x+
U_t\right)\cosh v \sinh v  \label{214b}
\end{eqnarray}
which are nonzero in general.
The expansion $\bar{\Theta}$ is given by the expression
\be
\bar{\Theta} = e^{-U}\left[ \left(v_t + U_x\right) \sinh v + \left(v_x
+ U_t\right)\cosh v \right] \label{215}
\ee
The shear tensor components have the form
\begin{eqnarray} \label{216}
\bar{\sigma}_{00} &=& \frac{1}{3} \left[(2\sinh^2v - 1)e^U + e^{-U}
\right]\left[(v_t + U_x)\sinh v + (v_x + U_t)\cosh v\right]
\label{216a} \\ \n \\
\bar{\sigma}_{01} &=& -\frac{2}{3}e^{U}\sinh v \cosh v
\left[(v_t + U_x)\sinh v + (v_x + U_t)\cosh v\right]
\label{216b} \\ \n \\
\bar{\sigma}_{11} &=& \frac{1}{3}  \left[(2\cosh^2 v +1)e^{U} -
e^{-U}\right]
\left[(v_t + U_x)\sinh v + (v_x + U_t)\cosh v\right]
\label{216c} \\ \n \\
\bar{\sigma}_{22} &=&  -\frac{1}{3} e^{2\nu - U}
\left[(v_t + U_x)\sinh v + (v_x + U_t)\cosh v\right] \n \\ \n \\
&=& \bar{\sigma}_{33}
\label{216d}
\end{eqnarray}
which do not vanish in general.  From (\ref{214})--(\ref{216}) we
observe that  the fluid congruences of the
conformally related line element (\ref{2}) are accelerating, expanding
and shearing.

\section{Reduction of the Field Equations}

The most general functional form admitted by the equations (\ref{212a})--(\ref{212d}) is
\be
e^{-U} = f(t,x) + h(y,z)   \label{217}
\ee
where $f$ and $h$ are arbitrary.
The remaining field equations (\ref{212e})--(\ref{212j}),
respectively, then assume the following form
\begin{eqnarray} \label{218}
2(f+ h) f_{tx} &=& -(\mu + p) \cosh v \sinh v   \label{218a}
\\ \n \\ h_{yz} &=& \nu_z h_y + \nu_y h_z \label{218b} \\ \n \\
-3\left( f^2_t - f^2_x \right) - 2(f+h)f_{xx} - 2k(f+h)^2&&  \n \\
\mbox{} + e^{-2\nu} \left[3 \left(h^2_y + h^2_z \right)
-2(f+h) \left( h_{yy} + h_{zz}\right) \right] &=& -(\mu +p) \cosh^2 v +p
\label{218c} \\ \n \\
-3\left( f^2_t - f^2_x \right) + 2(f+h)f_{tt} - 2k(f+h)^2 && \n \\
\mbox{} + e^{-2\nu} \left[3 \left(h^2_y + h^2_z \right)
-2(f+h) \left( h_{yy} + h_{zz}\right) \right] &=& (\mu +p) \sinh^2 v +p
\label{218d} \\ \n \\
-3\left( f^2_t - f^2_x \right) + 2(f+h)\left( f_{tt} - f_{xx}\right) &&
 \n \\ 
 \mbox{}+ e^{-2\nu} \left[3 \left(h^2_y + h^2_z \right)
-2(f+h) \left( \nu_yh_y - \nu_zh_z + h_{zz}\right) \right] &=&
p \label{218e} \\ \n \\
-3\left( f^2_t - f^2_x \right) + 2(f+h)\left( f_{tt} - f_{xx}\right) &&
 \n \\ 
\mbox{} + e^{-2\nu} \left[3 \left(h^2_y + h^2_z \right)
-2(f+h) \left( \nu_zh_z - \nu_yh_y + h_{yy}\right) \right] &=&
p \label{218f}
\end{eqnarray}
where the variable $U$ has been replaced by $f$ and $h$ via (\ref{217}).

\vspace{5mm}

The dynamical quantities have the following forms:\beq
p&=&-3\left(f^2_t - f^2_x\right) + 2(f+h)\left(f_{tt} - f_{xx}\right) \n
\\   &&\mbox{}+ e^{-2\nu}\left( 3\left(h^2_y + h^2_z\right) -
(f+h)\left(h_{yy} + h_{zz}\right) \right) \label{2192}
\eeq
for the pressure $p$ and
\be \mu = 3\left( f^2_t - f^2_x \right) + 4k(f+h)^2 - 3e^{-2\nu} \left(
(h^2_y + h^2_z) - (f+h)(h_{yy} + h_{zz}) \right) \label{220}
\ee
for the energy density $\mu$. We now seek an expression for the quantity
$v(t,x)$.
This is accomplished by first substituting (\ref{2192}) in
(\ref{218c}) for $p$ (but we retain the $p\cosh^2v$ term) to give
\be
(\mu + p)\cosh^2 v = 2(f+h)f_{tt} + 2k(f+h)^2 +
e^{-2\nu}(f+h)\left(h_{yy} + h_{zz}\right) \label{2193}
\ee
On substituting (\ref{2192}) into (\ref{218d}) for $p$ (but we retain
the $p\sinh^2v$ term) we get \be
(\mu +p)\sinh^2 v = 2(f+h)f_{xx} - 2k(f+h)^2 - e^{-2\nu}(f+h)\left(
h_{yy} + h_{zz}\right) \label{2194}
\ee
Dividing (\ref{2194}) by (\ref{2193}) yields
\be
\tanh ^2 v = \frac{2f_{xx} - 2k(f+h) - e^{-2\nu}(h_{yy}
+ h_{zz})} {2f_{tt} + 2k(f+h) + e^{-2\nu}(h_{yy} + h_{zz})} \label{221}
\ee
Another expression for $\tanh v$ may be obtained by
 dividing  (\ref{2194}) with
$(\ref{218a})$ to give
\[
\tanh^2v = \frac{\left( 2f_{xx} - 2k(f+h) - e^{-2\nu}(h_{yy} + h_{zz})
\right)^2}{4f^2_{tx}}
\]
Comparing  this equation with $(\ref{221})$ yields the differential
equation \beq
f^2_{tx} &=&
\left( 2f_{xx} - 2k(f+h) - e^{-2\nu}(h_{yy} + h_{zz})
\right)\n \\
&&\times \left( 2f_{tt} + 2k(f+h) + e^{-2\nu}(h_{yy} + h_{zz})\right)
\label{222}
\eeq
On rearranging (\ref{221}) we obtain
\[
\frac{2f_{xx} - 2f_{tt} \tanh^2v}{\tanh^2v + 1} - 2kf = 2kh +
e^{-2\nu} \left( h_{yy} + h_{zz} \right)
\]
from which it is clear that the left hand side is a function of
$t$ and $x$ whereas the right hand side is a function of $y$ and
$z$. This implies that the variables separate and  we can put \beq
\n \frac{2f_{xx} - 2f_{tt} \tanh^2v}{\tanh^2v + 1} - 2kf &=& 2kh +
e^{-2\nu} \left( h_{yy} + h_{zz} \right) \\ \n &=& \alpha \n \eeq
where $\alpha$ is a  constant. We thus derive the expression \be
h_{yy} + h_{zz} = e^{2\nu} ( \alpha - 2kh ) \label{223} \ee
Equation (\ref{223}) further simplifies the system
(\ref{218})--(\ref{218f}).

From the above analysis it is clear that the field equations
(\ref{218})--(\ref{218f}) can be expressed in a simpler form. The
resulting system is given by
\begin{eqnarray} \label{224}
\mu &=& 3(f^2_t - f^2_x) + (f+h)(4kf - 2kh + 3\alpha) -
3e^{-2\nu} \left( h^2_y + h^2_z \right) \label{224a}\\ \n \\
p &=&
- 3(f^2_t - f^2_x) + (f+h)\left(2f_{tt} - 2f_{xx} + 2kh -
\alpha\right)\n \\
&&\mbox{}+3e^{-2\nu} \left( h^2_y + h^2_z \right) \label{224b}\\ \n \\
\tanh^2v &=& \frac{2f_{xx} - 2kf - \alpha}{2f_{tt} + 2kf + \alpha}
\label{224c} \\ \n \\
f^2_{tx} &=& \frac{1}{4} \left(2f_{xx} - 2kf - \alpha\right)
\left(2f_{tt} + 2kf + \alpha \right) \label{224d} \\ \n \\
h_{yz} &=& \nu_{z} h_{y} + \nu_{y} h_z \label{224e} \\ \n \\
h_{yy} - h_{zz} &=& 2 \nu_y h_y + 2 \nu_z h_z \label{224f}
\end{eqnarray}
The system (\ref{224})--(\ref{224f}), subject to condition (\ref{2'}), viz.
$\nu_{yy} + \nu_{zz} = -2ke^{2\nu}$, must be
solved in order to generate a conformally related  perfect fluid model.

Castejon--Amenedo and Coley have considered the case $h=
\mbox{constant}$; this constant may be effectively absorbed into $f$
without any loss of generality and we can consequently set
\[
h = 0
\]
With this value of $h$ the Einstein field equations
(\ref{224})--(\ref{224f}) reduce to \beq \label{225} p &=& -3\left(
f^2_t - f^2_x \right) + 2f(f_{tt} - f_{xx})
\label{225a} \\ \n \\
\mu &=& 3\left( f^2_t - f^2_x \right) + 4kf^2 \label{225b} \\ \n \\
v&=& \tanh^{-1} \left( \frac{kf - f_{xx}}{f_{tx}} \right) \label{225c}
\\ \n \\
f^2_{tx} &=& \left( f_{tt} + kf \right) \left( f_{xx} - kf \right)
\label{225d}
\eeq
which is in agreement with the equations used by Castejon--Amenedo and
Coley (1992). Note that (\ref{225})--(\ref{225d}) corresponds to $h=\alpha=0$.  It is possible in (\ref{224})--(\ref{224f}) to have $\alpha=0$ with $h\neq0$.  This naturally leads to two categories of exact solutions that we now present.

\section{An Extension of the Castejon--Amenedo and Coley solutions: $h=0$}

This category of solutions corresponds to $h=\alpha=0$.
In order to generate a solution to the system (\ref{225})--(\ref{225d}) it
is sufficient to obtain a form for $f = f(t,x)$; in other words
(\ref{225d}) must be integrated. To obtain a solution
Castejon--Amenedo and Coley have assumed that the variables
separate and set \be f(t,x) = F(t) G(x) \ee which yields the
differential equations
\beq
F \ddot{F} - \gamma \dot{F}^2 + kF^2 &=& 0 \label{new1} \\
G G'' - \frac{1}{\gamma} G'^2 - kG^2 &=& 0, \label{new2}
\eeq  where $\gamma$ is a constant.

\subsection{CASE  I:  $\gamma = 1$}

It is easy to solve equations (\ref{new1})--(\ref{new2}) in general. As a result, the system (\ref{225})--(\ref{225c})
reduces to 
\beq \label{228}
\mu &=&
e^{k(x^2 - t^2) + 2k_1t + 2k_2 x}
\left[3k^2(t^2-x^2) - 6k(k_1t + k_2x) +
3(k^2_1 - k^2_2) + 4k\right] \n \\ \label{228a} \\ \n \\
p&=& e^{k(x^2 - t^2) + 2k_1t + 2k_2 x}
\left[-k^2(t^2-x^2) +2k(k_1t + k_2x) -  (k^2_1
- k^2_2) - 4k\right] \n \\ \label{228b} \\ \n \\
\tanh v &=& \frac{kx + k_2}{kt - k_1}. \label{228c} 
\eeq
This general solution contains the solution of Castejon--Amenedo and Coley  when we set
\[
k_1 = k_2 = 0
\]
in (\ref{228})--(\ref{228c}). For this choice of constants we
obtain \beq \label{2281} \mu &=& e^{k(x^2 - t^2)}
\left[3k^2(t^2-x^2)
+ 4k\right]  \label{2281a} \\ \n \\
p&=& e^{k(x^2 - t^2) }
\left[k^2(x^2 - t^2)
- 4k\right]  \label{2281b} \\ \n \\
\tanh v &=& \frac{x }{t} \label{2281c}  \eeq which is the
exact solution of Castejon--Amenedo and Coley (1992). Consequently
we have extended their solution (\ref{2281})--(\ref{2281c}) to
the more general class (\ref{228})--(\ref{228c}).

  For the exact solution (\ref{228})--(\ref{228c}) we obtain
\beq \n
\mu + p &=& 2
e^{k(x^2 - t^2) + 2k_1t + 2k_2 x}
\left[k^2(t^2-x^2) - 2k(k_1t + k_2x) +\left(k^2_1 - k^2_2\right)\right]
            \n \\ \n \\
\mu + 3p &=& -8k
e^{k(x^2 - t^2) + 2k_1t + 2k_2 x}
           \n \\ \n \\
\mu - p &=&
4e^{k(x^2 - t^2) + 2k_1t + 2k_2 x}
\left[k^2(t^2-x^2) - 2k(k_1t + k_2x) +\left(k^2_1 - k^2_2\right)+ 2k
\right]
                        \n
\eeq
and it is possible to study the weak, dominant and strong energy
conditions. We note that the appearance of the constants $k_1$ and $k_2$
allows for a wider range of behaviour for our class of solutions than is
the case with the Castejon--Amenedo and Coley (1992) exact solution.
For this model $\bar{\dot{u}}^a = 0$ so that the field is
nonaccelerating;
however both $\bar{\sigma}_{ab}$ and $\bar{\Theta}$ are nonzero which
implies that the gravitational  field is shearing and expanding.

\subsection{CASE II: $\gamma \neq 1$}

The case with $\gamma \neq 1$ in (\ref{225}) is of interest in
 generating cosmological models
for investigating  a wider range of physical behaviour.
The transformations
$$F = u_1^{\frac{1}{1-\gamma}}, \qquad G=u_2^{\frac{\gamma}{\gamma -
1}}$$ reduce equations (\ref{new1})--(\ref{new2}) to
 \beq \n
\ddot{u}_1 + k(1-\gamma) u_1 &=& 0 \n \\ \n \\
u''_2 + k\frac{(1 - \gamma)}{\gamma} u_2 &=& 0 \n
\eeq
which are linear in $u_1$ and $u_2$ respectively. The solutions of this system
 of differential equations are easily obtained in terms of
elementary functions and
consequently we do not present their explicit forms here.

\section{Group Invariant Solutions: $h\neq0$}

This category of solutions corresponds to $h\neq0$ with $\alpha=0$.  It is possible to solve the more general system (\ref{224})--(\ref{224f}) with $h\neq0$.  If we take $h=h(y)$, then (\ref{224e}) implies that $\nu=\nu(y)$.  Consequently, the condition (\ref{2'}) becomes the ordinary differential equation
\be \nu_{yy} = - 2 k e^{2\nu}. \label{alpha} \ee
Equation (\ref{224f}) reduces to
\be h_y = C e^{2\nu}, \label{beta} \ee
where $C$ is a constant of integration.  The general solution to (\ref{alpha}) and (\ref{beta}) is given by
\beq
e^{2\nu} &=& -\frac{A}{k} \mbox{cosech}^2\left( \sqrt{2A}y + B \right) \label{gamma}
 \\ \n \\
h(y) &=& \frac{C\sqrt{A}}{\sqrt{2} k} \coth \left( \sqrt{2A} y +
B\right) + D,  \label{delta}
\eeq
where $A, B$ and $D$ are constants of integration.  In order to complete the solution we need to solve (\ref{224d}) and obtain $f$ as this
immediately yields expressions for $p$, $\mu$ and $v$.
We now focus our attention on (\ref{224d}) with $\alpha=0$, i.e.
\be
f^2_{tx} = \left( f_{tt} + kf \right) \left( f_{xx} - kf \right) \label{Delta} 
\ee
 with the objective of
generating solutions in a systematic manner. We achieve this by
 invoking the method of Lie group analysis of differential equations (Olver 1993).
The Lie symmetry generators of (\ref{Delta}) are found with the help
of the computer package
{\tt PROGRAM LIE} (Head 1993). The infinitesimal Lie symmetry generators
 for the partial differential equation (\ref{Delta}) are
\beq \label{229}
{Z}_1 &=&\frac{\partial}{\partial t} \label{229a} \\ \n \\
{Z}_2 &=&\frac{\partial}{\partial x} \label{229b} \\ \n \\
{Z}_3 &=&x\frac{\partial}{\partial t} + t \frac{\partial}{\partial
x} \label{229c} \\ \n \\
{Z}_4 &=&f \frac{\partial}{\partial f} \label{229d} \eeq It is
clear that these symmetries form the Lie algebra is $A_{3,4}
\oplus A_{1}$ (Patera and Winternitz 1977) with basis
given by \beq
G_1 &=&Z_1 + Z_2  \\ \n \\
G_2 &=&Z_1 - Z_2  \\ \n \\
G_3 &=&Z_3  \\ \n \\
G_4 &=&Z_4.
\eeq
It is this basis that we will use in our subsequent analysis. Before proceeding further, we observe that (\ref{Delta}) is invariant under the following discrete transformations:
\be
f\rightarrow -f \label{r1a}
\ee
\be
t\rightarrow -t \label{r1b}
\ee
\mbox{and}
\be
x\rightarrow -x \label{r1c}
\ee
Taking these reflections into account, we have
\be
G_{4}\rightarrow -G_{4} \label{r2d}
\ee
\be
G_{2}\rightarrow -G_{1} \label{r2b}
\ee
\mbox{and}
\be
G_{3}\rightarrow -G_{3} \label{r2c}
\ee
We also note that $x\rightarrow -x$ will make $G_{1} = G_{2}$.

In order to obtain group invariant solutions of (\ref{Delta}) (with $f =
f(t,x)$ explicitly)
we only need consider the following symmetry combinations (Msomi 2003):
\beq
G_{1} &=& \frac{\partial }{\partial t}+\frac{\partial }{\partial x} \label{e1} \\
G_{4} &=& {f}\frac{\partial }{\partial f} \label{e2}\\
G_{1}+G_{4} &=& \frac{\partial }{\partial t} + \frac{\partial }{\partial x} + {f}\frac{\partial }{\partial f} \label{e3}\\
G_{3}+\beta G_{4} &=& {x}\frac{\partial }{\partial t} +
{t}\frac{\partial }{\partial x} + \beta {f}\frac{\partial }{\partial f} \label{e5}\\
G_{1}+G_{2}+\beta G_{4} &=& 2\frac{\partial }{\partial t} + \beta
{f}\frac{\partial }{\partial f} \label{e6} \eeq where $\beta$ is a
real arbitrary parameter. All other solutions of (\ref{Delta})
obtained via linear combinations of the point symmetries
(\ref{229a})--(\ref{229d}) can be obtained from the solutions we
report here. We consider each in turn.

\subsection{Invariance under $G_{1}$} Using $G_{1}$ we determine the invariants from the
invariant surface condition \beq \frac{dt}{1} &=
\frac{dx}{1}=\frac{df}{0} \eeq which yields \beq
{y} &=&x-t\\
{f} &=&{V}.
\eeq
Now by using the above transformation, in equation (\ref{Delta}),
the partial differential equation is reduced to
\beq
k^{2}V^{2} &=&0 \label{2.6}
\eeq
\mbox{ie.}
\beq
{V} &=&{0}.
\eeq
Thus only trivial travelling wave solutions are possible.
\subsection{Invariance under $G_{4}$}
In this case the system has invariants given by
\beq
{y} &=&t\\
{V} &=&x.
\eeq
From the above transformation, we cannot reduce (\ref{Delta}). There is no ordinary differential equation to solve.

\subsection{Invariance under $G_{1}+G_{4}$}
Here, the invariants are
\beq
{y} &=&x-t \label{dys}\\
{f} &=&{Ve^{t}}. \label{san} \eeq The partial differential
equation is reduced by the transformation to the form \beq
\left(k\left(1+k\right)V^{2}+V^{2}_{y}-V\left(2KV_{y}+V_{yy}\right)\right)
&=&0 \label{2.7} \eeq which is linearisable and has solution \be
\log V =
\frac{(1+k)}{2}y-\frac{(1+k)}{4k}+\frac{C_{0}}{2k}+C_{1}e^{-2ky},
\ee where $C_0$ and $C_1$ are arbitrary constants, whence \beq f
&=&\exp\left({\frac{(1+k)}{2}x +
\frac{(1-k)}{2}t-\frac{(1+k)}{4k}+\frac{C_{0}}{2k}+C_{1}e^{-2k(x-t)}}\right)\label{0.11}
\eeq and so we have ``time--boosted'' travelling wave solutions.
With $h$ given by (\ref{delta}) and $f$ by (\ref{0.11}) it is possible to show that there are regions for which $\mu >
0$ and $p > 0$. This is a desirable feature because we expect that
barotropic matter in cosmological models should have positive
pressures and positive energy densities. 

\subsection{Invariance under $G_{3}+\beta G_{4}$} \label{special}
For this case the system has invariants given by
\beq
{y} &=&x^{2}-t^{2}\\
{f} &=&\left(x+t\right)^{\beta}{V} \eeq which reduces the partial
differential equation to \be
k^{2}V^{2}+4(V_{y}(1+\beta))^{2}+8yV_{y}V_{yy}-4VV_{y} k(1+\beta
)-4VV_{yy}ky +4VV_{yy} \beta(1-\beta) = 0. \label{2.16}
\ee This equation is an ordinary differential equation with the
single symmetry \beq Z_{1} &=&{V}\frac{\partial }{\partial V}.
\label{2.17} \eeq In order to solve the differential equation
(\ref{2.16}) with one symmetry, we first try to reduce the order
of this equation and see if the resulting equation is easily
integrated.

The differential equation (\ref{2.16}) has the following reduction variables from (\ref{2.17}):
\beq
r &=&y\\
q &=&\frac{V_y}{V}. \eeq Using these invariants in (\ref{2.16}), we have \be
\frac{dq}{dr} =
\frac{-k^{2}-4q^{2}(1+\beta)^{2}+4q(1+\beta)}{8pq-4qkp-4(-1+\beta)\beta)}-q^{2}.\label{0.2020}
\ee In this case, we find that the first order equation cannot be
easily integrated. We would have to resort to numerical solutions.
However, some special solutions can still be found as demonstrated
later.

\subsection{Invariance under $G_{1}+G_{2}+\beta G_{4}$}
In this final case  the invariants are
\beq
{y} &=&x\\
{f} &=&Ve^{\frac{t\beta}{2}}. \eeq The partial differential
equation is reduced by this transformation to the form
\be \left(  {\beta}^{2}+4\,k  
 \right)V V_{yy}  - {\beta}^{2}V_y^2- k\left({\beta}^{2}+4\,{k}\right) V^{2} = 0 
\label{error} \ee which has general solution  
\be V = {2}^{{\frac {{\beta}^{2}+4\,k}{4k}}} \left( - \frac{\left( {\it C_1}\,
\sin \psi -{\it 
C_2}\,\cos \psi 
 \right) ^{2}}{ \left( {\beta}^{2}+4\,k \right)}  \right) ^{{
\frac {{\beta}^{2}+4\,k}{8k}}},\ee where $C_1$ and $C_2$ are
arbitrary constants and
 \be \psi = \frac{2kx}{\sqrt{-\beta^2-4k}}. \ee Thus a solution to (\ref{Delta}) is given by
\be f = {2}^{{\frac {{\beta}^{2}+4\,k}{4k}}}e^{\frac{t\beta}{2}} \left( - \frac{\left( {\it C_1}\,
\sin \psi -{\it 
C_2}\,\cos\psi 
 \right) ^{2}}{ \left( {\beta}^{2}+4\,k \right)}  \right) ^{{
\frac {{\beta}^{2}+4\,k}{8k}}}. \label{0.3030} \ee

\section{A Particular Solution}
In order to investigate the physical properties of the spacetimes
obtained, we take a special case of the intractable result in  \S
\ref{special}, that of $\beta=0$.  In this case the new
independent variable is given by
 \be u = t^2 - x^2\ee and the
new functional form is
\be f(t,x) = f(u) = f(t^2 - x^2). \ee As a result,  (\ref{Delta}) becomes
\be
4u\frac{d^2f}{du^2} + 2\frac{df}{du} + kf = 0. \label{232}
\ee
 We now
define new variables $a$ and $q$ via \be f(u) = a (u) u^{1/4}
\qquad q = u^{1/2}
 \ee and obtain the equation
\be
q^2\frac{d^2a}{dq^2} + q\frac{da}{dq} + \left(kq^2 -
\left(\frac{1}{2}\right)^2 \right)a = 0. \label{235}
\ee
We now distinguish between the two cases $k > 0$ and $k < 0$.

For  $k> 0$, (\ref{235})
 can be simplified to
\be
w^2\frac{d^2a}{dw^2} + w\frac{da}{dw} + \left( w^2 -
\left(\frac{1}{2}\right)^2 \right)a = 0 \label{236}
\ee
by  the transformation $\sqrt{kq} = w$. Equation (\ref{236}) is
the Bessel equation of order $\frac{1}{2}$ and its solutions are the
linearly independent Bessel functions $J_{\frac{1}{2}}(w)$ and
$J_{-\frac{1}{2}}(w)$ which may be expressed in terms of elementary
trigonometric functions in the following way:
\beq \n
J_{\frac{1}{2}}(w) &=&\sqrt{\frac{2}{\pi w}} \sin w \n \\  \n \\
J_{-\frac{1}{2}}(w) &=&\sqrt{\frac{2}{\pi w}} \cos w \n.
\eeq
It is a pleasing feature of this model that (\ref{Delta}) admits
Bessel functions as solutions since many realistic phenomena are
governed by the Bessel equation. The solution to the differential
equation (\ref{Delta}), in the original variables $t$ and $x$,  is
\be
f(x,t) = A \sin \sqrt{k(t^2 - x^2)} + B \cos \sqrt{k(t^2 - x^2)}
\label{237}
\ee
where $A$ and $B$ are arbitrary constants.
For the solution (\ref{237})
the quantity $v$ is given by
\[
\tanh v = \frac{t}{x}
\]
so  that the exact  solution corresponding to  (\ref{237})  is not
conformally flat.

For $k < 0$, we make the substitution $k = - \beta^2$ followed by
$W = \beta z$. Equation (\ref{235}) then has the form
\[
W^2\frac{d^2a}{dW^2} + W\frac{da}{dW} - \left(W^2 +
\left(\frac{1}{2}\right)^2\right) = 0
\]
which is the modified Bessel equation of order $\frac{1}{2}$. This
differential equation admits the modified Bessel functions
\beq 
I_{\frac{1}{2}} &=&\sqrt{\frac{2}{\pi W}} \sinh W \\ \n \\
I_{-\frac{1}{2}} &=&\sqrt{\frac{2}{\pi W}} \cosh W 
\eeq
as linearly independent solutions. The solution to the field equation
(\ref{Delta}), in terms of the original variables $t$ and $x$,  may be written as
\be
f(x,t) = A \sinh \sqrt{k\left(t^2 - x^2\right)} + B \cosh \sqrt{k\left(t^2 -
x^2\right)}
\ee
analogous to (\ref{237}).

\section{Conclusion}
The notion of conformally mapping a given line element to a new metric, such as the conformally related metric (\ref{2}), is potentially a very fertile avenue in generating new solutions to the Einstein field equations.  However, the actual number of exact solutions found using this algorithm is very low as pointed out by Stephani {\it et al} (2003).  The Petrov type D spacetime turns out to be a rare metric which is amenable to this approach, and we have therefore focussed our attention on this line element (first considered by Castejon--Amenedo and Coley (1992)).  The Petrov type D model investigated is physically well behaved as there is a barotropic equation of state, the weak and strong energy conditions are satisfied and it is amendable to a simple two perfect fluid interpretation.

In order to demonstrate a new solution we had to find new functions $h$ and $f$ which define the conformal function $U$ in (\ref{2}).  Two categories of solutions arise naturally in the analysis.  In the first category ($h=0$) we found a general class of solutions in terms of elementary functions.  This contains the solutions of Castejon--Amenedo and Coley (1992) as special cases.  In the second category ($h\neq0$) we
invoked the method of
Lie group
analysis to solve a master  field equation which allowed us to obtain various exact solutions
for the metric under consideration. Four cases arise depending on the Lie symmetry generators (we ignore the non-reducible case). Three
of the cases were readily solvable, in general, in terms of elementary functions. However, we were only able to provide special solutions in the remaining case.

This was analysed further for physical plausibility by considering
a particular solution. We demonstrated that in this case that the
solution can be expressed in terms of Bessel and modified Bessel
functions. The positivity of pressure and energy density in this case ensured its relevance for the description of some cosmological processes. All solutions found via the Lie method could be used in both $h=0$ and $h\neq0$ cases. This
treatment demonstrates the importance of the method of Lie group
analysis in seeking solutions to the Einstein field equations.

\section*{Acknowledgments}

SDM and KSG thank the University of KwaZulu-Natal and the National Research Foundation of South Africa
for ongoing support.

\section*{References}
\begin{description}
\item{} Castejon--Amenedo J and Coley A A 1992 {\em {Class.
Quantum Grav.}} {\bf 9} 2203 \item{} Choquet--Bruhat Y, 
DeWitt--Morette C and Dillard--Bleick M 1982 {\em Analysis,
manifolds and physics} (Amsterdam: North Holland) \item{} Coley A
A and Tupper B O J 1990a {\em Gen. Rel. Grav. }{\bf 22} 241
\item{} Coley A A and Tupper B O J 1990b {\em Class. Quantum Grav.
}{\bf 7} 1961 \item{} Coley A A and Tupper B O J 1990c {\em Class.
Quantum Grav. }{\bf 7} 2195
\item{} Defrise-Carter L 1975 {\em
Commun. Math. Phys.} {\bf 40} 273 \item{} Dyer C C, McVittie G C
and Oates L M 1987 {\em Gen. Rel. Grav. } {\bf 19} 887 \item{}
Head A K 1993 {\em Comp. Phys. Comm.} {\bf 71} 241 \item{} Kramer
D 1990 {\em Gen. Rel. Grav. } {\bf 22} 1157 \item{} Kramer D and
Carot J 1991 {\em J. Math. Phys.} {\bf 32} 1857
   \item{} Maartens R and Maharaj S D 1986
{\em Class. Quantum Grav.} {\bf 3} 1005 \item{} Maartens R and
Maharaj S D 1991 {\em Class. Quantum Grav.} {\bf 8} 503 \item{}
Maartens R, Maharaj S D and Tupper B O J 1995 {\em Class. Quantum
Grav.} {\bf 12} 2577 \item{} Maartens R, Maharaj S D and Tupper B
O J 1996 {\em Class. Quantum Grav.} {\bf 13} 317 \item{} Maartens
R and Mellin C M 1996 {\em Class. Quantum Grav.} {\bf 13} 1571
\item{} Maharaj S D and Leach P G L 1996 {\em{J. Math. Phys.}}
{\bf 37} 430 \item{} Maharaj S D, Leach P G L and Maartens R 1991
{\em{Gen. Rel. Grav.}} {\bf 23} 261 \item{} Mars M and Senovilla J
M M 1993 {\em Class. Quantum Grav.} {\bf 10} 1633 \item{} Mars M
and Senovilla J M M 1994 {\em Class. Quantum Grav.} {\bf 11} 3049
\item Moodley M 1992 {\em  Conformal symmetries: Solutions in two
classes of cosmological models} (MSc dissertation, University of
Natal, Durban,  South Africa) \item Moopanar S 1993 {\em Conformal
motions in general relativity} (PhD dissertation, University of
Natal, Durban,  South Africa)
 \item{} Msomi A M 2003 {\em An
Application of Optimal Subgroups to Partial Differential
Equations} (MSc dissertation, University of Natal, Durban,  South
Africa) \item{} Olver P J 1993 {\em Applications of Lie Groups to
Differential Equations} (New York: Springer--Verlag) \item{}
Patera J and Winternitz P 1977 {\it J. Math. Phys.}  {\bf 18} 1449
\item[] Penrose R 1999  {\it Chaos Solitons Fractals} {\bf 10} 581
-- 611
\item{} Stephani H, Kramer D, Maccallum M, Hoenselaers C and Herlt E 2003 {\it Exact Solutions to Einstein's Field Equations} (Cambridge: Cambridge University Press)
\end{description}
\end{document}